# Do Spherical Polyelectrolyte Brushes Interdigitate?

A.V. Korobko,[1] W. Jesse,[1] S.U. Egelhaaf,[2] A. Lapp,[3] and J.R.C. van der Maarel[1,4,*]

[1]*Leiden Institute of Chemistry, Leiden University, 2300 RA Leiden, the Netherlands*
[2]*School of Physics and School of Chemistry, The University of Edinburgh, Edinburgh EH9 3JZ, United Kingdom*
[3]*Laboratoire Léon Brillouin, CEA/CNRS, 91191 Gif-sur-Yvette Cedex, France*
[4]*Department of Physics, National University of Singapore, Singapore 119260, Singapore*

The structure of spherical micelles of the diblock copolymer poly(styrene-*block*-acrylic acid) in water was investigated with small angle neutron scattering (SANS) and contrast matching. We have monitored inter-micelle correlation and the extension of the polyelectrolyte chains in the coronal layer through the overlap concentration. Irrespective of ionic strength, the corona shrinks with increasing packing fraction. Furthermore, at high charge and minimal screening conditions, the corona layers interpenetrate once the volume fraction exceeds the critical value 0.53±0.02.

Applications of polyelectrolyte brushes are numerous: from stabilization of colloidal suspensions, through control of flow behavior, to cell adhesion and drug delivery [1-3]. Polyelectrolyte brushes can be prepared by anchoring ionic chains at an interface [4,5] or one can use the principle of self-assembly of diblock copolymers [6-12]. They can be classified according to their morphology, including planar, cylindrical, and spherical brushes. The key concept in understanding of their functioning is the structure of the brush in terms of polymer density and counterion binding. Contrary to neutral brushes, stretching of the polyelectrolyte brush is primarily effected by the osmotic pressure exerted by counterions adsorbed in the layer, rather than the repulsion between monomers.

Spherical micelles formed by the aggregation of diblock copolymers typically consist of a neutral core surrounded by a polyelectrolyte coronal brush. For individual micelles, the corona size and its relation to charge, screening, and counterion distribution have been investigated [6-11]. The main results are osmotic star-branched polyelectrolyte behavior, full corona chain stretching at high charge and minimal screening conditions, similar counterion and corona segment density profiles, and charge annealing effects toward the outer corona region at low degrees of ionization.

Despite the considerable body of work, not much is known about the organization among micelles and how the structure of the corona changes when the micelles interact. In particular, the extent to which the coronal layers contract or interdigitate upon an increase in concentration is an open question [7]. Concentrated polyelectrolyte copolymer systems have vast technological potential due to providing control of, *e.g.*, gelation, lubrication, and flow behavior [1,2]. It is our contention that the behavior of interacting polyelectrolyte brushes, including interdigitation, is important in understanding the fluid properties.

Here, we report small angle neutron scattering (SANS) experiments on a model system of spherical micelles up to concentrations where the coronas have to shrink and/or interpenetrate in order to accommodate the micelles in the increasingly crowded volume (the functionality is fixed due to a glassy core). We focus on the predicted contraction of the corona before overlap, possible interdigitation at high packing fraction, and the relation with charge and electrostatic screening [13,14]. The core and corona structure factors, as obtained from contrast matching in water, are interpreted in terms of core size, inter-micelle correlation, and statistical properties of the corona-forming segments. Comparison of the micelle diameter from the form factor analysis with the diameter from the center of mass structure factor and with the average inter-micelle distance will then show the extent to which the coronal layers interpenetrate.

We express the core or corona (*i*) structure factor as

$$S_i(q) = P_i(q) S_{cm}(q) \quad (1)$$

with form factor $P_i(q)$ and micelle center of mass structure factor $S_{cm}(q)$ [15]. The core can be described by a homogeneous sphere with diameter $D_{core}$. For the corona, we adopt an algebraic radial density profile $\rho_{corona} \sim r^{-\alpha}$, $D_{core} < 2r < D_{mic}$ with outer micelle diameter $D_{mic}$. The value of $\alpha$ is determined by the chain statistics [13]. At high charge and minimal screening conditions, the chains are almost fully stretched and $\alpha = 2$. In excess salt, the additional screening of Coulomb interaction results in a radial decay similar to neutral star-branched polymers with $\alpha = 4/3$ [16]. At low degree of ionization, the scaling exponent takes the value 8/3 due to charge annealing effects. In the long wavelength limit ($q \to 0$), the form factors are normalized to the number of copolymers per micelle, *i.e.* the aggregation number $N_{ag}$.

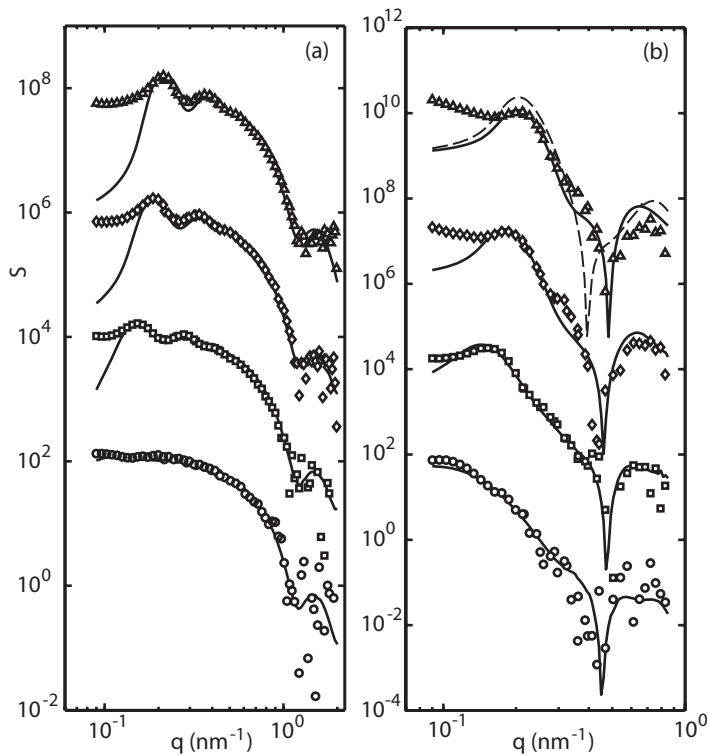

FIG. 1. Core PS (a) and corona PA (b) structure factor versus momentum transfer for fully charged PS-b-PA micelles without added salt. The copolymer concentration is 44 ($\triangle$), 30 ($\diamond$), 17 ($\square$), and 4.4 ($\circ$) g/l from top to bottom. The data are shifted along the y-axis. The solid lines represent the model calculations. The dashed line denotes the force fit of the corona structure factor with the hard sphere diameter for the highest concentration.

For polyelectrolyte copolymer micelles, an analytic expression for the center of mass structure factor is not available. We have analyzed the data with a hard sphere potential and the Percus-Yevick approximation for the closure relation [17]. The fit parameters are the micelle density $\rho$ and hard sphere diameter $D_{hs}$. The hard sphere diameter should be interpreted as an effective diameter; its value could be smaller than the outer micelle diameter if interpenetration occurs. Furthermore, it is known that for soft objects the hard sphere potential does not correctly predict the relative amplitudes of the primary and higher order correlation peaks [18]. We have also tested a repulsive screened Coulomb potential [19]. However, the effect of electrostatic interaction among the micelles was found to be modest, which is attributed to the fact that almost all neutralizing counterions are confined in the coronal layer [10].

We studied micelles formed by poly(styrene-*block*-acrylic acid) [PS-*b*-PA] (Polymer Source Inc.) with degrees of polymerization 20 and 85 of the PS and PA blocks, respectively. At ambient temperature, the PS core is in a glassy state, which results in micelles with fixed core size and functionality. The PA corona charge is pH dependent and can be varied between almost zero and full (100%) charge where every monomer carries an ionized group. Six sets of solutions with 100, 50, and 10% corona charge were prepared: 3 sets without added salt, in another 3 the salt (KBr) concentration is 1.0 M (100 and 50% charge) or 0.04 M (10% charge). Each set was prepared with 4 copolymer concentrations ranging from the dilute to the dense regime, where the coronas should interpenetrate if they do not shrink [20]. Furthermore, we applied contrast variation with 4 solvent compositions: 0%, 29% (PS-matched), 70% (PA-matched), and 100% $D_2O$. SANS was measured at ambient temperature with the D22 and PAXY diffractometers situated on the cold sources of the Institute Laue-Langevin and Laboratoire Léon Brillouin, respectively. A wavelength of 0.8 nm with a 10% spread was selected. We obtained the core and corona structure factors by a simultaneous 2-parameter fit to the data from the 4 solvent compositions [9-11]. As an example, the results pertaining to the fully charged micelles without added salt are displayed in Fig. 1. The corresponding micelle center of mass structure factor is shown in Fig. 2 (for all, but the lowest concentration).

At the lowest micelle concentration, inter-micelle interference is insignificant and the core and corona structure factors can directly be compared with the relevant form factors. With increasing concentration and minimal screening conditions, a primary and higher order correlation peaks emerge. The position of the primary peak scales with the copolymer concentration $C_{pol}$ according to $C_{pol}^{1/3}$, which is characteristic for micelles with fixed aggregation number and local spherical symmetry. There is no major change in the high $q$ behavior of the corona structure factor with increasing packing fraction. This shows that the chains remain almost fully stretched and $\alpha = 2$. The lines in Fig. 1 represent the model calculations with form factor parameters $D_{core} = 9$ nm and $D_{mic}$ displayed in Fig. 3 (the parameters pertaining to the fit of the center of mass structure factor, $D_{hs}$ and $\rho$, are discussed below).

With excess salt, inter-micelle interference is largely suppressed and the corona structure factors are compared with the form factor calculated with $\alpha = 4/3$ (100 and 50% charge) or $8/3$ (10% charge). The fitted micelle diameters are also displayed in Fig. 3.

With added salt and/or at low degree of ionization, the coronal layers are less extended. The ionic strength and charge dependencies of the micelle diameter agree with our previous results obtained for more diluted samples [9-11]. With increasing packing fraction, the diameter of the micelles, as obtained from the form factor analysis, decreases. However, the extent to which the coronal layers shrink is modest and similar under all conditions. The gradual decrease in size is due to interaction among micelles, increased counterion adsorption, and/or Donnan salt partitioning between the coronal layer and the supporting medium [5,13].

From the normalization of the structure factors, an aggregation number $N_{ag}$ around 100 is derived, irrespective of charge, copolymer concentration, and ionic strength.

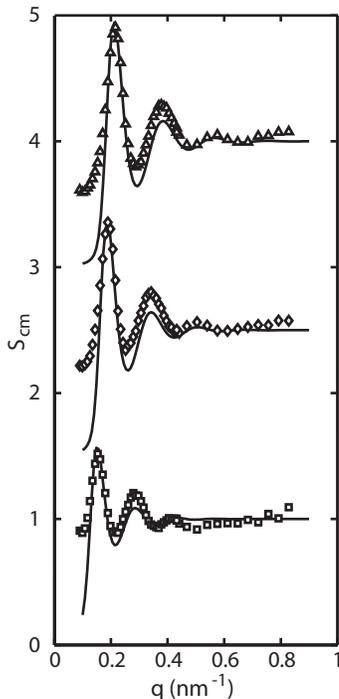

FIG. 2. Center of mass structure factor versus momentum transfer (full charge, without added salt). The concentrations are as in Fig. 1. The data are shifted along the y-axis. The curves represent the hard sphere structure factor.

Inter-micelle interference is more clearly demonstrated in Fig. 2, where the core structure factor has been divided by the core form factor (full charge and no added salt). Although the center of mass structure factor could also be derived from the corona structure factor, we have chosen to use the core structure factor because the core form factor shows a smooth and moderate variation in the relevant $q$-range (even so, consistency with the corona structure factor is illustrated in Fig. 1b). The intensity of the correlation peaks first increases and eventually levels off with increasing packing fraction, which shows the progressive and saturating ordering of the micelles. Notice that for the present volume fractions the position of the primary peak is mainly determined by density, whereas the respective positions of the higher order correlation peaks are most sensitive to the value of the hard sphere diameter. The lines in Fig. 2 represent the hard sphere solution structure factor convoluted with the instrument resolution with fitted micelle densities and hard sphere diameters displayed in Fig. 3. The hard sphere model is capable of predicting the positions of the primary and higher order peaks. Furthermore, the ratio of the fitted micelle densities and known copolymer concentrations provides an alternative way to obtain the aggregation number, $N_{ag} = 98\pm10$. This value is in perfect agreement with the one obtained from the normalization of the structure factors.

Clear deviations between the experimental data and the hard sphere prediction are observed in the low $q$-range. Furthermore, the model underestimates the intensity of the second order peak with respect to the primary one. We have checked that a repulsive, screened Coulomb potential does not improve the fit, nor does it significantly influence the peak positions for reasonable values of the micelle charge (the net micelle charge is small, because almost all counterions are confined in the coronal layer [10]). The failure in predicting the relative amplitude of the higher order peak is probably related to the softness of the micelles; similar behavior has been reported for interpenetrating neutral polymer stars [18]. The deviations observed in the low $q$-range might be due to long-range inhomogeneity in density, the formation of aggregates, and/or stickiness between the micelles. Although the latter phenomenon is likely, we have refrained from interpreting our data with more elaborate models, such as the sticky hard sphere model [21].

The hard sphere diameters were derived for salt-free micelles only, because in the presence of excess salt inter-micelle interference is effectively suppressed. For the less concentrated, 17 g/l solutions, the hard sphere diameters equal the micelle diameters derived from the form factor analysis (Fig. 3). This supports the applicability of the hard sphere interaction model in order to extract the effective hard sphere diameters. At higher packing fractions and for the 50 and 100% charged micelles in particular, the effective hard sphere diameters are significantly smaller than the outer micelle diameters ($D_{mic}$ and $D_{hs}$ can be estimated within 3 and 2% error margins, respectively). As shown in Fig. 1, a force fit with only a hard sphere diameter is not acceptable. We take the difference as a measure of the extent to which the corona layers interpenetrate. Accordingly, the 100 and 50% charged, salt-free micelles interpenetrate around 17 g/l; for the smaller 10% charged micelles this happens at a higher concentration, say 25 g/l.

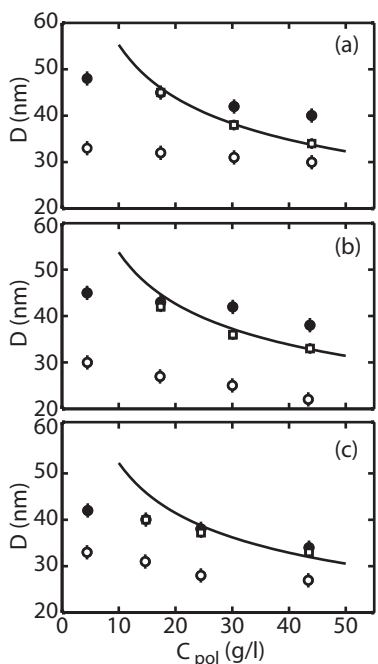

FIG. 3. Concentration dependence of the diameter of 100 (a), 50 (b), and 10% (c) charged PS-*b*-PA micelles: (●), $D_{mic}$ without added salt; (○), $D_{mic}$ in 1.0 M (100 and 50%) or 0.04 M (10% charge) KBr. The hard-sphere diameter of salt-free micelles $D_{hs}$ is denoted by (□). The lines represent $\rho^{-1/3}$, *i.e.* the average inter-micelle distance.

Fig. 3 also displays the average distance between the micelles $\rho^{-1/3}$. Once the micelles interpenetrate, the effective hard sphere diameter equals $\rho^{-1/3}$. Based on the optimized densities and hard sphere diameters, effective micelle volume fractions are calculated. For interpenetrating micelles, the effective volume fraction is found to be constant within experimental accuracy and takes the value 0.53±0.02. Notice that, although this volume fraction corresponds with closely packed, simple cubic order, the center of mass structure factor remains liquid-like and no long-range order in the diffraction patterns is observed. Interdigitation thus occurs when the volume fraction exceeds the critical value 0.53. For higher copolymer concentration, this value is effectively preserved by interpenetration of the coronal layers.

We conclude that with increasing packing fraction and minimal screening conditions the micelles shrink and the coronal layers eventually interdigitate. This effect is most pronounced for higher corona charge. In the presence of excess salt, the coronas also contract, but interpenetration does not occur in the present concentration range.

We acknowledge the Institute Laue-Langevin and Laboratoire Léon Brillouin in providing the neutron research facilities. This research has been supported by the Netherlands Organization for Scientific Research and the European Community's HPRI program.